# Undergraduate Signal Processing Laboratories for the Android Operating System

Suhas Ranganath, JJ Thiagarajan, KN Ramamurthy, Shuang Hu,
Mahesh Banavar, and Andreas Spanias
{srangan8,jjayaram,knatesan,Shuang.Hu.1,mbanavar,spanias}@asu.edu

**Abstract**

We present a DSP simulation environment that will enable students to perform laboratory exercises using Android mobile devices and tablets. Due to the pervasive nature of the mobile technology, education applications designed for mobile devices have the potential to stimulate student interest in addition to offering convenient access and interaction capabilities. This paper describes a portable signal processing laboratory for the Android platform. This software is intended to be an educational tool for students and instructors in DSP, and signals and systems courses. The development of Android JDSP (A-JDSP) is carried out using the Android SDK, which is a Java-based open source development platform. The proposed application contains basic DSP functions for convolution, sampling, FFT, filtering and frequency domain analysis, with a convenient graphical user interface. A description of the architecture, functions and planned assessments are presented in this paper.

**Introduction**

Mobile technologies have grown rapidly in recent years and play a significant role in modern day computing. The pervasiveness of mobile devices opens up new avenues for developing applications in education, entertainment and personal communications. Understanding the effectiveness of smartphones and tablets in classroom instruction have been a subject of considerable research in recent years. The advantages of handheld devices over personal computers in K-12 education have been investigated[1]. The study has found that the easy accessibility and maneuverability of handheld devices lead to an increase in student interest. By incorporating mobile technologies into mathematics and applied mathematics courses, it has been shown that smartphones can broaden the scope and effectiveness of technical education in classrooms [2].

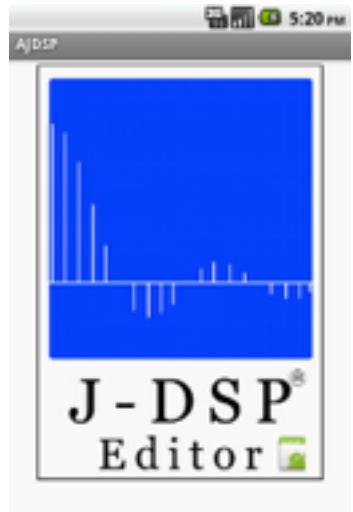

Fig 1: Splash screen of the A- JDSP Android application

Designing interactive applications to complement traditional teaching methods in STEM education has also been of considerable interest. The role of interactive learning in knowledge dissemination and acquisition has been discussed and it has been found to assist in the development of cognitive skills[3]. It has been showed learning potential is enhanced when education tools that possess a higher degree of interactivity are employed[4]. Software applications that incorporate visual components in learning, in order to simplify the understanding of complex theoretical concepts, have been also been developed[5-9]. These applications are generally characterized by rich user interaction and ease of accessibility.

Modern mobile phones and tablets possess abundant memory and powerful processors, in addition to providing highly interactive interfaces. These features enable the design of applications that require intensive calculations to be supported on mobile devices. In particular, Android operating system based smartphones and tablets have large user base and sophisticated hardware configurations. Though several applications catering to elementary school education have been developed for Android devices, not much effort has been undertaken towards building DSP simulation applications[10]. In this paper, we propose a mobile based application that will enable students to perform Digital Signal Processing laboratories on their smartphone devices (Figure 1).

In order to enable students to perform DSP labs over the Internet, the authors developed J-DSP, a visual programming environment [11-12]. J-DSP was designed as a zero footprint, standalone Java applet that can run directly on a browser. Several interactive laboratories have been developed and assessed in undergraduate courses. In addition to containing basic signal processing functions such as sampling, convolution, digital filter design and spectral analysis, J-DSP is also supported by several toolboxes. An iOS version of the software has also been developed and presented [13-15]. Here, we describe an Android based graphical application, A-JDSP, for signal processing simulation. The proposed tool has the potential to enhance DSP education by supporting both educators and students alike to teach and learn digital signal processing.

The rest of the paper is organized as follows. We review related work in Section 2 and present the architecture of the proposed application in Section 3. In Section 4 we describe some of the functionalities of the software. We describe planned assessment strategies for the proposed application in Section 5. The concluding remarks and possible directions of extending this work are discussed in Section 6.

**Related Work**

Commercial packages such as MATLAB[16] and LabVIEW[17] are commonly used in signal processing research and application development. J-DSP, a web-based graphical DSP simulation package, was proposed as a non-commercial alternative for performing laboratories in undergraduate courses [3]. Though J-DSP is a light-weight application, running J-DSP over the web on mobile devices can be data-intensive. Hence, executing simulations directly on the mobile device is a suitable alternative.

A mobile application that supports functions pertinent to different areas in electrical engineering, such as circuit theory, control systems and DSP has been reported [18]. However, it does not contain a comprehensive set of functions to simulate several DSP systems. In addition to this, a mobile interface for the MATLAB package has been released[19]. However, this requires an active version of MATLAB on a remote machine and a high speed internet connection to access the remote machine from the mobile device. In order to circumvent these problems, i-JDSP, an iOS version of the J-DSP software was proposed [13-15]. It implements DSP functions and algorithms optimized for mobile devices, thereby removing the need for internet connectivity. Our work builds upon J-DSP [11-12] and the iOS version of J-DSP [13-15], and proposes to build an application for the Android operating system. Presently, to the best of our knowledge, there are no freely available Android applications that focus on signal processing education.

**Architecture**

The proposed application is implemented using Android-SDK [22], which is a Java based development framework. The user interfaces are implemented using XML as it is well suited for Android development. The architecture of the proposed system is illustrated in Figure 2. It has five main components: (i) User Interfaces, (ii) Part Object, (iii) Part Calculator, (iv) Part View, and (v) Parts Controller. The role of each of them is described below in detail.

The blocks in A-JDSP can be accessed through a function palette (user interface) and each block is associated with a view using which the function properties can be modified. The user interfaces obtain the user input data and pass them to the Part Object. Furthermore, every block has a separate Calculator function to perform the mathematical and signal processing algorithms. The Part Calculator uses the data from the input pins of the block, implements the relevant algorithms and updates the output pins.

Figure 2. Architecture of A- JDSP.

All the configuration information, such as the pin specifications, the part name and location of the block is contained in the Part Object class. In addition, the Part Object can access the data from each of the input pins of the block. When the user adds a particular block in the simulation, an instance of the Part Object class is created and is stored by a list object in the Parts Controller. The Parts Controller is an interface between the Part Object and the Part View. One of the main functions of Parts Controller is supervising block creation. The process of block creation by the Parts Controller can be described as follows: The block is configured by the user through the user interface and the block data is passed to an instance of the Part Object class. The Part Object then sends the block configuration information through the Parts Controller to the Part View, which finally renders the block.

The Part View is the main graphical interface of the application. This displays the blocks and connections on the screen. It contains functionalities for selecting, moving and deleting blocks. Examples of block diagrams in the A-JDSP application for different simulations are illustrated in Figure 3(a), Figure 4(a) and Figure 5(a) respectively.

**Functionalities**

In this section, we describe some of the DSP functionalities that have been developed as part of A-JDSP.

*Android based Signal Generator block*

This generates the various input signals necessary for A-JDSP simulations. In addition to deterministic signals such as square, triangular and sinusoids; random signals from Gaussian Rayleigh and Uniform distributions can be generated. The signal related parameters such as signal frequency, time shift, mean and variance can be set through the user interface.

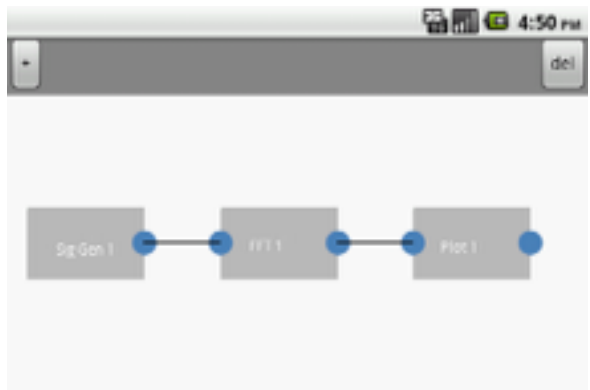 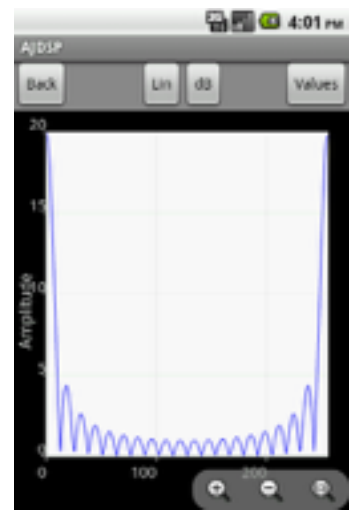

(a) (b)

Figure 3. (a) Block Diagram of FFT (b) Plot of FFT of a Rectangular Signal.

*Android based FFT block*

The algorithm for the Fast Fourier transform has been implemented in this block. The discrete Fourier transform is calculated using the Cooley-Tukey decimation-in-time FFT algorithm [21]. Different lengths of FFT can be chosen by the user. Figure 3 shows an example simulation in

A- JDSP for computing FFT of a rectangular signal. The plot view in Figure 3 can be selected by pressing the Plot function and the user can navigate back to the block diagram by pressing the Back button.

*Android based Filtering block*

Filtering is one of the basic algorithms of signal processing. Several functionalities have been provided to help students to understand concepts in filtering. Windowing functions have been implemented for both time and frequency domain signals. The set of windowing functions supported in A-JDSP include Rectangular, Bartlett, Hamming, Hanning and Kaiser windows. Functions for designing FIR filters such as the Kaiser and Parks McClellan, and IIR filters such as the Butterworth, Chebyshev (1 and 2), and Blackmann have been implemented. Figure 4(a) and 4(b) shows the block diagram and the frequency response for a low-pass Kaiser Filter while Figure 5(a) and 5(b) illustrates the same for a high-pass Butterworth filter. The frequency response plots can be displayed by placing the *Freq Resp* function in the simulation. This function accepts filter co-efficients as the input and plots the phase and magnitude responses of the filter.

*Android based Plot functions*

Plotting functions help the students to visualize signal processing concepts. A-JDSP supports three types of plot functions: (a) basic plot, (b) frequency response plot, and (c) pole-zero plot. In order to implement the plot functions, we used the *achartengine library* for Android [20]. Note that, both linear and logarithmic plot options have been implemented.

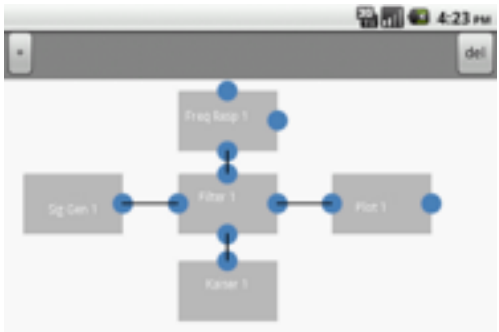 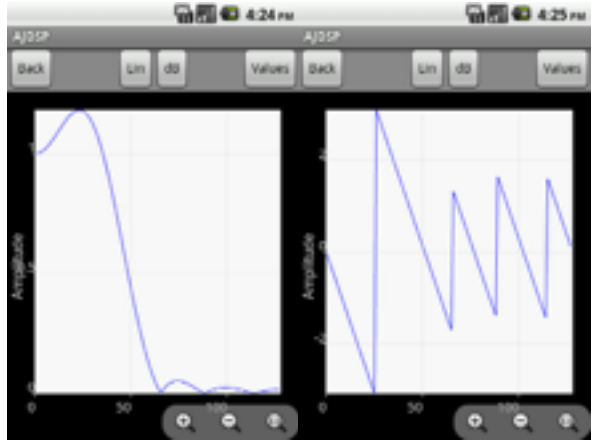

(a)                                                        (b)

Figure 4. (a) Setup for Kaiser filter simulation (b) Magnitude and Phase Responses for a Low-Pass Kaiser Filter.

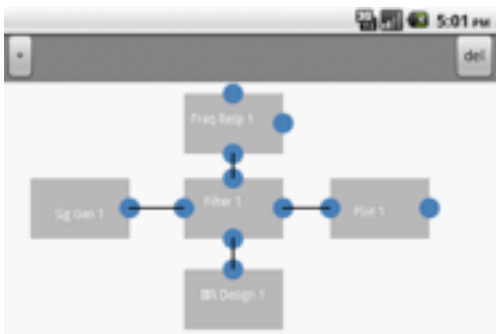 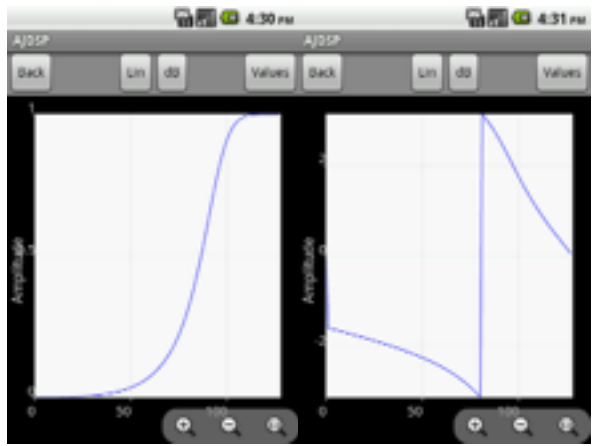

(a)                                                        (b)

Figure 5. (a) Setup for IIR filter simulation (b) Magnitude and Phase Responses for a High-Pass Butterworth Filter.

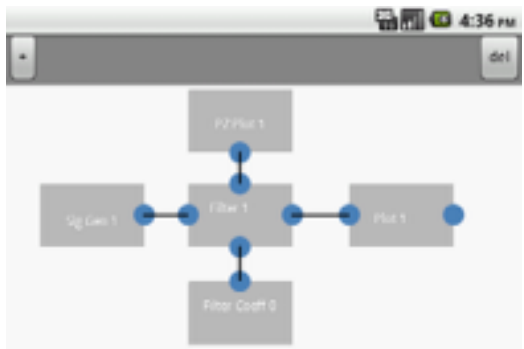
(a)

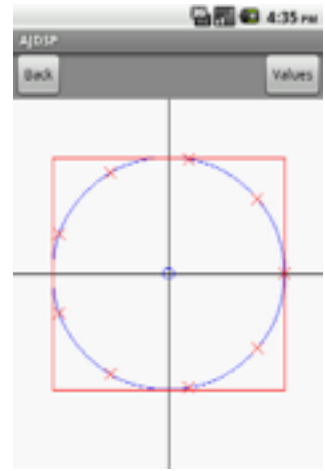
(b)

Figure 6. (a)Setup for Pole-Zero demonstration (b) Pole-zero diagram for $H(z) = \frac{1}{1-z^{-10}}$

*Android based PZ Plot block*

Functionality for visualizing the pole zero diagrams of filters has been provided. This block computes and displays the pole-zero plots taking the filter coefficients of the filter as the input. Figure 6 illustrates the pole zero plot of an IIR Butterworth filter for the transfer function $H(z) = \frac{1}{1-z^{-10}}$ as generated by the PZ Plot block of A-JDSP.

**Assessments**

We are currently designing a set of laboratory exercises for A- JDSP to be performed by students of an undergraduate signal processing class at Arizona State University during Spring 2012. Furthermore, we will generate assessment results based on the student evaluation of the application. The goal of this evaluation is to identify the impact of employing mobile devices to perform DSP simulations. The assessments will determine if the A-JDSP framework was interactive and interesting for students to perform different simulations. Furthermore, general assessments about the aesthetics and the usability can be used to obtain an overall subjective opinion about the application. The pedagogy adopted for the use of A- JDSP will include the following: (a) lecture on the pertinent signal processing concepts, (b) a pre-quiz on the concepts involved in the laboratory exercise, (c) a simulation exercise using A-JDSP, (d) post-quiz to test student understanding of the concepts. We list a part of our assessment questionnaire below.

- Did the contents of the A-JDSP exercises improve your understanding of the concepts of filter design?
- Did the exercises help you understand the effects of causality?
- If a student colleague is having difficulty understanding filter design, would you recommend the A-JDSP convolution demonstration?
- Is it practical to perform DSP exercises and simulations on an Android Device?
- Does using your Android phone, does that give you a more compelling reason to finish your lab exercises than using a computer based tool?
- How easy is it to navigate through the application to create a simulation?
- How long did it take for you to learn filter design A-JDSP?

Addressing the issues identified based on the student evaluation will enable us to provide prescriptive recommendations concerning strengths, replication, and sustainability of this setup.

**Conclusions**

In this paper, we described a graphical DSP programming application for Android OS based phones and tablets. The proposed application has been developed using the Java based Android SDK and is compatible with all Android devices. We described the architecture of the application and presented the DSP functions in A-JDSP. The current set of functions in the application will enable students to perform simulation exercises on convolution, Fourier analysis and filter design. The interface is highly interactive and the block diagrams can be constructed using a simple space-and-route procedure. Finally, we described our planned assessments in order to understand the impact of this application in performing DSP laboratories.

**Acknowledgements**

This project is supported in part by NSF award 0817596, the SenSIP center, and Sprint Communications.